\documentclass[12pt]{iopart}
\usepackage{placeins}
\usepackage{hyperref}
\usepackage{braket}
\usepackage{subcaption}
\usepackage{graphicx}
\makeatletter\let\@captype\relax\makeatother
\usepackage[font=small,labelfont=bf]{caption}

\usepackage{iopams}

\usepackage{iopams}  
\begin{document}

%\title[Thermodynamic and statistical entropies of quantum systems]{Thermodynamic and statistical entropies of quantum systems}
\title[Nonextensive Thermodynamics of the Morse Oscillator: Signature and Solid State Application]{Nonextensive Thermodynamics of the Morse Oscillator: Signature and Solid State Application}
\author{Arpita Goswami$^\dagger$}
\address{Department of Physics, Indian Institute of Technology Tirupati, India, 517619}
\ead{$^\dagger$ph23d007@iittp.ac.in}

\vspace{10pt}

\begin{abstract} 
In this work, we present a detailed thermodynamic analysis of a bound quantum system: the Morse oscillator within the framework of Tsallis nonextensive statistics. Using the property of the bound spectrum (upper bound) of the Morse potential, limited by the bond dissociation energy, we analytically derive the generalized partition function. We present results for both the high- and low-temperature limits. We propose the effective number of accessible states as a measure of nonextensivity. The calculation shows that the nonextensive framework further restricts the number of accessible states. We also derive the generalized internal energy and entropy and examine their dependence on temperature and the nonextensivity parameter \( q \). Numerical results confirm the strong effect of nonextensive behavior in the low-temperature regime (precisely low to moderate temperature), where the ratio of generalized internal energy and internal energy calculated from the Boltzmann Gibbs (BG) formula develops a nontrivial dip structure for \( q < 1 \). Moreover, the generalized specific heat shows the Schottky-type anomaly. We extend our study by deriving the specific heat of solids with BG and Tsallis statistics using the anharmonic energy levels of the Morse oscillator. This study suggests that the Morse oscillator is a solvable and physically meaningful testing ground for exploring the thermodynamics of quantum systems driven by nonextensive statistics, with implications for the vibrational properties of the non-equilibrium molecular thermodynamics (especially diatomic molecules).

\vspace{0.35cm}
\end{abstract}

\tableofcontents

\section{Introduction} \label{sec1}

Boltzmann–Gibbs (BG) statistical mechanics has long provided a robust framework for describing equilibrium thermodynamics in a wide variety of physical systems. However, several classes of systems—those exhibiting long-range interactions, memory effects, fractal geometries, or non-equilibrium stationary states—often fall outside its descriptive reach. To address these regions of complexity, several generalized statistical frameworks have been developed, among them the nonextensive formalism proposed by Tsallis~\cite{Tsallis1988} has been proven particularly fruitful and gained scholarly attention over the past few decades.

The formalism introduces a deformation parameter \( q \in \mathbb{R} \), which characterizes the degree of nonextensivity in the system. In the limit \( q \to 1 \), the standard BG formalism is recovered, while deviations emerge for \( q \neq 1 \). Over the past decades, this formalism has been successfully applied to a wide range of problems, including the Boltzmann H-theorem~\cite{PhysRevLett.86.2938}, Ehrenfest relations~\cite{PLASTINO1993177}, von Neumann entropy~\cite{Qi2021}, quantum statistics~\cite{DEPPMAN2021166}, fluctuation-dissipation theorems, and generalized Langevin and Fokker–Planck dynamics~\cite{PLASTINO1995347}. Applications have also been extended to magnetic systems~\cite{PhysRevB.55.5611}, infinite-range spin models~\cite{NOBRE1995337}, Planck radiation law generalizations~\cite{TIRNAKLI1997657}, and systems with underlying fractal or chaotic structures~\cite{El-Nabulsi12122022, HUANG20251354}.

Despite its extensive study, most applications of Tsallis statistics on oscillator systems are mainly restricted to idealized models, such as classical and quantum harmonic oscillators~\cite{CHAKRABARTI20084589, PLASTINO2017196, Ishihara2024, Ito1989, gomez2025, ANDRADE1991285}—which do not fully capture the anharmonic nature of real physical systems. Realistic systems rarely show perfect harmonicity; instead, interactions involve nonlinear restoring forces, energy cutoffs arising from the dissociation thresholds, or may include resistive forces. One such natural example is the Morse oscillator, which provides a more precise description of molecular vibrations, particularly in diatomic systems. It features an inherently anharmonic potential with a finite number of bound states, bounded above by the bond dissociation energy.

Theoretical studies of the Morse oscillator were initiated primarily to address its thermal properties, in the BG framework~\cite{AMOREBIETA1981530}, using approaches such as Euler–Maclaurin expansions~\cite{Boumali2018}, integral equation techniques~\cite{AlRaeei2022}, and numerical methods. However, its study under nonextensive statistics remains largely unexplored. This study provides a systematic investigation of the Morse oscillator within the Tsallis framework that combines both analytical asymptotic analysis and numerical evaluation of thermodynamic quantities. 
\par
Moreover, recent advances in the trapping and cooling of atomic gases have allowed precise control over microscopic degrees of freedom and made the low-dimensional molecular potentials increasingly relevant for experimental realizations. The Morse oscillator, with its finite bound states and anharmonic energy spacing, offers a more realistic alternative to the harmonic approximation for modeling interatomic bonds in ultracold regimes. Within this context, the application of Tsallis nonextensive statistics becomes particularly valuable: it accounts for bounded spectra, compact phase-space structures, and long-range interactions that emerge in finite-temperature, trapped cold atom ensembles. Similar thermodynamic anomalies and compact configurational supports have been observed in hybrid atom-ion platforms \cite{Dutta_2020}, nanophotonic traps that exhibit cooperative decay \cite{PhysRevX.14.031004}, and cold molecular ions confined in optical lattices \cite{deiss2024molecular}. These experimental platforms provide promising grounds where Tsallis-based models, especially ones incorporating anharmonic potentials like Morse, may offer meaningful physical insight.
Motivated by these insights, we aim to explore how nonextensivity affects the thermodynamics of anharmonic quantum oscillators with bounded spectra. Our study not only advances the field of atomic physics, but by using the Morse oscillator to get the specific heat of solids, opens a connection with condensed matter theory. To the best of our knowledge, no previous study has systematically explored both the molecular thermodynamics and specific heat of solids with the Morse oscillator after combining nonextensivity, where the Morse oscillator is used to study solid-state physics.

In this work, we derive analytical expressions for the Tsallis-deformed partition function \( Z_q(T) \) of the Morse oscillator in both the high- and low-temperature regimes. We then compute generalized thermodynamic quantities such as internal energy \( U_q \), the effective number of contributing states \( n_{\mathrm{eff}} \), and entropy \( S_q \). Our results reveal significant deviations from the behavior of BG at intermediate temperatures and in small \( q \), where the compact support of the \( q \)-exponential becomes dominant. At high temperatures, all quantities converge smoothly to their BG limits. In the end, we also derive the specific heat of solids using a perturbative approach in both BG and Tsallis statistics. As the system Morse oscillator has an upper bound to the accessible state in the spectra, so we will only restrict our study to the $q<1$ limit since it is known that the $q<1$ limit provides a subextensive (assigns more statistical weight to the low lying states and less to the higher energy states) picture of the system and $q>1$ gives a superextensive nature (exact opposite to the subextensive case). As the upper bound of the system is fixed, the superextensive case ($q>1$) becomes less important for the system. 

The study is organized as follows: we provide the Tsallis formalism and its application to the Morse oscillator in Sec.~\ref{sec2}.  We derive analytical expressions of different thermodynamic quantities in various temperature regimes in Sec.~\ref{sec3}.  We provide numerical results that confirm and generalize our analytical predictions in Sec.~\ref{sec:anharmonicity} and Sec.~\ref{sec4}. Then, in Sec.~\ref{new_sec}, we give a brief idea of how the effect of the deformation parameter can be physically realized.  Lastly, in Sec.~\ref{secapp}, we derive the specific heat expression with both the BG and Tsallis statistics in both high and low temperature limits. Sec.~\ref{sec5} provides a summary of our results and outlines future research directions.

\section{Tsallis Formalism and Model System} \label{sec2}

\subsection{Preliminaries of the formalism}

We begin by reviewing the essential definitions and mathematical tools from the Tsallis nonextensive statistical mechanics.

The generalized entropy, known as Tsallis entropy~\cite{q_val}, is defined as:
\begin{equation}
    S_q = k_B \ln_q W,
\end{equation}
where \( q \in \mathbb{R}_+ \) is the nonextensivity parameter, \( k_B \) is the Boltzmann constant, and \( W \) is the number of microstates corresponding to a given macrostate. The parameter \( q \) quantifies the degree of nonextensivity; the standard BG entropy is recovered in the limit \( q \to 1 \)~\cite{Tsallis1988}. 

To construct a consistent thermodynamic framework, the standard logarithmic and exponential functions are replaced by their \( q \)-deformed counterparts:
\begin{eqnarray}
    \ln_q(x) &= \frac{x^{1 - q} - 1}{1 - q}, \quad
    \exp_q(x) &= \left[1 + (1 - q)x\right]_+^{1/(1 - q)},
\end{eqnarray}
where \( [x]_+ = \max\{0, x\} \). These deformed functions reduce to the natural logarithm and the exponential in the BG limit \( q \to 1 \).

\subsection{Tsallis canonical distribution}

In the canonical ensemble, the Tsallis probability distribution is given by:
\begin{equation}
    p_n = \frac{ \left[1 - (1 - q)\beta E_n \right]^{1/(1 - q)} }{Z_q}, \quad \mathrm{for} \quad 1 - (1 - q)\beta E_n > 0,
\end{equation}
where \( \beta = 1/(k_B T) \) is the inverse temperature and \( E_n \) is the energy of the \( n \)-th microstate. The generalized partition function is defined as
\begin{equation}
    Z_q = \sum_{n = 0}^{n_{\mathrm{max}}} \left[1 - (1 - q)\beta E_n \right]^{1/(1 - q)}.
    \label{eq:Zq}
\end{equation}

For \( q < 1 \), the \( q \)-exponential has a compact support, meaning only energy levels that satisfy \( (1 - q)\beta E_n < 1 \) contribute. However, for the Morse oscillator, the number of bound states is finite and is naturally bounded above by \( n_{\mathrm{max}} \), so the support cutoff is automatically satisfied and no artificial truncation is needed.

\subsection{Thermodynamic quantities}

Given the probability distribution \( p_n \), the internal energy in the Tsallis framework is:
\begin{equation}
    U_q = \sum_{n = 0}^{n_{\mathrm{max}}} E_n \, p_n^q.
    \label{eq:Uq}
\end{equation}
This should be compared with the BG internal energy:
\begin{equation}
    U_{\mathrm{BG}} = \sum_{n = 0}^{n_{\mathrm{max}}} E_n \, \frac{e^{-\beta E_n}}{Z_1}, \quad \mathrm{with} \quad Z_1 = \sum_{n = 0}^{n_{\mathrm{max}}} e^{-\beta E_n}.
\end{equation}

To quantify the number of states that contribute significantly in a given \( T \) and \( q \), we define the \textit{effective number of accessible states}:
\begin{equation}
    n_{\mathrm{eff}} = \left( \sum_{n = 0}^{n_{\mathrm{max}}} p_n^q \right)^{-1}.
    \label{eq:Neff}
\end{equation}
In the BG limit \( q \to 1 \), this reduces to the entropy-weighted effective count of microstates.

We will evaluate all of the above thermodynamic quantities both analytically and numerically in subsequent sections.

\subsection{Model system: the Morse oscillator}

The Morse potential models the vibrational interaction between atoms in a diatomic molecule and is given by:
\begin{equation}
    V(x) = D_e \left(1 - e^{-a(x - x_0)}\right)^2,
\end{equation}
where \( D_e \) is the bond dissociation energy, \( a \) characterizes the potential width, and \( x_0 \) is the equilibrium bond length. The corresponding Hamiltonian is:
\begin{equation}
    H = \frac{P^2}{2m} + V(x),
\end{equation}
and the bound-state energy levels are:
\begin{equation}
    E_n = \hbar \omega_0 \left(n + \frac{1}{2}\right) - \hbar \omega_0 x_e \left(n + \frac{1}{2}\right)^2, \qquad n = 0, 1, \dots, n_{\mathrm{max}}.
    \label{eq:En}
\end{equation}

Here, \( \omega_0 \) is the characteristic vibrational frequency whose value is $\omega_0=a\sqrt{\frac{2D_e}{m}}$ with m to be the mass, and \( x_e = \hbar \omega_0 / (4D_e) \) is the anharmonicity constant. The total number of bound states is finite and is given by:
\begin{equation}
    n_{\mathrm{max}} = \left[\frac{ \sqrt{2 D_e m} }{ \hbar a } - \frac{1}{2} \right] \simeq \frac{1}{2x_e}-\frac12,
\end{equation}
where the floor function ensures the largest integer is less than the cutoff value.

The presence of a finite energy spectrum makes the Morse oscillator particularly suitable for applying Tsallis statistics. Unlike systems with unbounded energy levels, the natural cutoff in \( E_n \) ensures that the \( q \)-exponential remains well-defined across the temperature range and avoids the divergences often encountered in Tsallis-based models with infinite spectra.

\section{Analytical Results} \label{sec3}

In this section, we derive analytical expressions for the Tsallis partition function \( Z_q(T) \), the effective number of accessible states \( n_{\mathrm{eff}} \), and the internal energy \( U_q \) in various temperature regimes. We focus on both high-temperature (continuum) and low-temperature (few-level) approximations for \( Z_q(T) \) calculation.

\subsection{High-temperature approximation for \texorpdfstring{$Z_q(T)$}{Zq(T)}}

At high temperatures (\( k_BT \simeq D_e \)), the spacing between energy levels becomes small and the sum in Eq.~\ref{eq:Zq} can be approximated by an integral. Using the substitution \( x = n + \frac{1}{2} \), and inserting the energy expression from Eq.~\ref{eq:En}, the partition function becomes:
\begin{equation}
Z_q(T) \simeq \int_{1/2}^{x_{\mathrm{max}}} \left[1 - (1 - q) \beta (A x - B x^2) \right]^{\frac{1}{1 - q}} \, dx,
\end{equation}
where \( A = \hbar \omega_0 \) and \( B = \hbar \omega_0 x_e \).

For small \( \beta \) (i.e., high temperature), we expand the integrand in a power series:
\begin{eqnarray}
\left[1 - (1 - q) \beta (Ax - Bx^2)\right]^{\frac{1}{1 - q}} &\simeq&
1 - \beta (Ax - Bx^2) \nonumber \\
&& + \frac{( q)\beta^2}{2} (Ax - Bx^2)^2.
\end{eqnarray}
Higher-order terms in \( \beta \) are neglected. Integrating term-by-term from \( x = 1/2 \) to \( x_{\mathrm{max}} \), we obtain:
\begin{eqnarray}
&Z_q(T)_{\mathrm{H}} &\simeq n_{\mathrm{max}}  
- \beta \left[ \frac{A ((n_{\mathrm{max}}+\frac12)^2-\frac14)}{2} - \frac{B ((n_{\mathrm{max}}+\frac12)^3-\frac 18)}{3} \right] \nonumber \\
&& + \frac{(q)}{2} \beta^2 \left[ \frac{A^2 ((n_{\mathrm{max}}+\frac12)^3-\frac{1}{8})}{3}-\frac{2AB ((n_{\mathrm{max}}+\frac12)^4-\frac{1}{16})}{4}  \right] \nonumber \\
&& 
+ \frac{(q)}{2} \beta^2\left[
\frac{B^2 ((n_{\mathrm{max}}+\frac12)^5-\frac{1}{32})}{5} \right].
\end{eqnarray}
This shows that at high temperatures, \( Z_q(T)_H \) becomes nearly independent of \( q \), and the Tsallis and BG frameworks coincide. Moreover, expectedly, all the thermodynamic quantities at high temperature will be q-independent. As the results confirm that there is no q-dependence on the high temperature limit, for the analysis of thermodynamic quantities, we will mainly focus on the moderate temperature regime where the nonextensive effects are strong.

\subsection{Moderate to Low-temperature approximation for \texorpdfstring{$Z_q(T)$}{Zq(T)}}

At extremely low temperatures, only a few low-lying energy levels contribute significantly. So, the sum can be performed exactly. However, we approximate $Z_q(T)$ for a temperature region that is moderate to low (ML), more precisely, the cases in which neither the high-temperature series expansion nor the exact low-temperature summation over a few degrees of freedom can work. We are in a temperature region for which $ k_BT > \Delta E,$ and $D_e > k_BT,$ where $\Delta E=\hbar\omega_0$ and $D_e$ are the energy gap between two successive energy states and the bond dissociation energy, respectively. By considering the thermodynamic limit, the sum can be effectively taken as an integral:
\begin{equation}
Z_q(T) \simeq \int_{1/2}^{x_{\mathrm{max}}} \left[1 - (1 - q) \beta (Ax - Bx^2) \right]^{\frac{1}{1 - q}} dx.
\end{equation}

Defining the following shifted variable:
\begin{equation}
z = \sqrt{(1 - q) \beta B} \left(x - \frac{A}{2B} \right), \qquad R^2 = 1 - \frac{(1 - q)\beta A^2}{4B},
\end{equation}
we can recast the integral as follows:
\begin{equation}
Z_q(T) \simeq \frac{1}{\sqrt{(1 - q)\beta B}} \int_{0}^{z_{\mathrm{max}}} \left[R^2 + z^2 \right]^{\frac{1}{1 - q}} dz.
\end{equation}
At this stage, we will do another variable transformation. We can replace $z=R~\tan{\theta}$, so, $dz=R~\sec^2{\theta}~d\theta$ (with $R^2 >0$). We can see that at the above maintained temperature limit, $ \beta B $ (as $\beta B=\frac{\hbar \omega_0 x_e}{k_B T}=\frac{x_e\Delta E}{k_B T}$, and $x_e<1$ ) is small, thus the upper cutoff limit corresponding to $n_{\mathrm{max}}$ can effectively be considered to be $\infty$ (if we keep $D_e$ to be very high and we are in the temperature limit $\hbar \omega_0 < k_B T < D_e$ then system has a large number of states to access thus $n_{\mathrm{max}} \to \infty$ effectively). Thus, the above integration takes the following form:
\begin{equation}
Z_q(T) \simeq \frac{R^{\frac{3 - q}{1 - q}}}{\sqrt{(1 - q)\beta B}} \int_{0}^{\frac \pi2} ({\sec{\theta}})^{\frac{2}{1 - q}} \sec^2{\theta} d\theta.
\end{equation}

Using the Euler-Beta function, the integral evaluates to:
\begin{equation} \label{ML_partition}
Z_q(T)_{\mathrm{ML}} = \frac{R^{\frac{3 - q}{1 - q}}}{2\sqrt{(1 - q)\beta B}} \mathrm{B\bigg(\frac12,\frac{q-3}{2(1-q)}\bigg)}.
\end{equation}
where $\mathrm{B}\bigg(\frac12, \frac{q-3}{2(1-q)}\bigg)$ is the beta function and can be converted to a gamma function with the relation \[\mathrm{B(n,m)=\frac{\Gamma{(m)}\Gamma{(n)}}{\Gamma{(m+n)}}}\] Substituting this into Eq.~\ref{ML_partition} we get the following result:
\begin{equation}
Z_q(T)_{\mathrm{ML}} = \frac{R^{\frac{3 - q}{1 - q}}}{2\sqrt{(1 - q)\beta B}} 
\frac{\Gamma(1/2)\Gamma\left(\frac{q - 3}{2(1 - q)}\right)}{\Gamma\left(-\frac{1}{1 - q}\right)}.
\end{equation}
This expression explicitly shows the dependence on \( q \), with the leading-order temperature scaling \( Z_q \sim 1/\sqrt{\beta} \).
However, because of the application of Tsallis statistics, we get the correction terms $h(q)=\frac{R^{\frac{3 - q}{1 - q}}}{2\sqrt{(1 - q)}} 
\frac{\Gamma(1/2)\Gamma\left(\frac{q - 3}{2(1 - q)}\right)}{\Gamma\left(-\frac{1}{1 - q}\right)}.$ Thus, the modified partition function under Tsallis statistics can be written as:
\begin{equation} \label{partition_analytics}
Z_q(T)_{\mathrm{ML}} = \frac{h(q)}{\sqrt{\beta B}}.
\end{equation} The expression of $h(q)$ shows that the leading order q dependence is proportional to $\frac{1}{\sqrt{1-q}}.$ This confirms that the deviation will be higher for $q$ values far from 1.
\subsection{Effective cut off states \texorpdfstring{$n_{\mathrm{eff}}$}{max}} In this section, our main goal is to find the effect of nonextensivity on the upper bound of the quantum state for the Morse oscillator. To do that, first let us take the generalized Boltzmann factor in the Tsallis statistics:
\begin{equation} \label{n_f}
    p_q(n) \propto \left[1 - (1 - q)\beta E_n\right]_+^{\frac{1}{1 - q}},
\end{equation}
where \( [x]_+ \) represents the nonnegative argument of the function. From Eq.~\ref{n_f}, we should have the following condition to obtain compact support of the generalized Boltzmann factor:
\begin{equation}
    1 - (1 - q)\beta E_n \geq 0.
\end{equation}

This leads us to a modified 'cutoff' condition on the energy:
\[
    E_n \leq \frac{1}{(1 - q)\beta}.
\]

To find the maximum allowed state \( n_{\mathrm{eff}} \), we substitute the expression for $E_n$ from Eq.~\ref{eq:En} and get the following equation:
\begin{equation}
    E_n = \hbar \omega_0 \left(n + \frac{1}{2}\right) - \hbar \omega_0 x_e \left(n + \frac{1}{2}\right)^2 \leq \frac{1}{(1 - q)\beta}.
\end{equation}

Let \( x = n + \frac{1}{2} \). Then we get the following:
\[
    E(x) = \hbar \omega_0 x (1 - x_e x) \leq \frac{1}{(1 - q)\beta}.
\]

The above can be written as the following quadratic inequality:
\begin{equation}
    x (1 - x_e x) \leq \frac{1}{\hbar \omega_0 (1 - q)\beta}.
\end{equation}

Let \( A_1 = x_e \), \( B_1 = \frac{1}{\hbar \omega_0 (1 - q)\beta} \). Then the inequality takes the form:
    \[x - A_1 x^2 \leq B_1.\]

This can be further rearranged as follows:
\[
    A_1 x^2 - x + B_1 \geq 0.
\]

Solving the quadratic equation, we get the following:
\begin{equation}
    x = \frac{1 \pm \sqrt{1 - 4A_1 B_1}}{2A_1}.
\end{equation}

We take the 'smaller root' (as the energy increases with \( x \)), and we define:
\begin{equation}
    x_{\mathrm{eff}} = \frac{1 - \sqrt{1 - 4 A_1B_1}}{2 A_1}=\frac{1}{2x_e}-\sqrt{\frac{1}{4x^2_e}-\frac{1}{x_e \hbar \omega_0(1-q)\beta}}.
\end{equation}

Then, the q-dependent modified expression of the maximum accessible state can be expressed as:
\[
    n_{\mathrm{eff}} = \left[ \frac{1}{2x_e}-\sqrt{\frac{1}{4x^2_e}-\frac{1}{x_e \hbar \omega_0(1-q)\beta}} - \frac{1}{2} \right].
\]
\begin{equation}\label{analytics_neff}
    n_{\mathrm{eff}} = \left[ n_{\mathrm{max}} -\sqrt{\frac{1}{4x^2_e}-\frac{1}{x_e \hbar \omega_0(1-q)\beta}} ~~\right].
\end{equation}
It should be noted that the square root is real only if:
\begin{equation} \label{T_scal}
    1 - 4 A_1B_1 \geq 0 \quad \Rightarrow \quad \beta \geq \frac{4x_e}{ \hbar \omega_0 (1 - q)}.
\end{equation}

This gives us the criteria or limit of the minimum inverse temperature ($\beta$) for which, in a given \( q < 1 \), the above derivation of $n_{\mathrm{eff}}$ is valid. The result suggests that within the above temperature limit and for $q<1$ for a given temperature, a small $q$ has a smaller $n_{\mathrm{eff}}$ value than the actual. However, with increasing q values, the number of accessible states also increases. This result suggests that due to the introduction of the Tsallis parameter q in the partition function, the effective number of bound states decreases. Eq.~\ref{T_scal} also shows that the temperature at which the deviation of the effective number of states ($n_{\mathrm{eff}}$) disappears scales as: $T \propto (1-q).$ This is the temperature at which the q-deformed statistics recover the BG statistics.

\subsection{Deformed internal energy \texorpdfstring{$U_q$}{Uq}}
In this section, we derive the generalized internal energy expression to see how the deviating (correction) term depends on the temperature and the q-values.
The internal energy is defined via Eq.~\ref{eq:Uq}, which in the continuum limit becomes:
\begin{equation}
U_q \simeq \frac{\int_0^{n_{\mathrm{max}}} E_n \left[\exp_q(-\beta E_n)\right]^q dn}
{\int_0^{n_{\mathrm{max}}} \left[\exp_q(-\beta E_n)\right]^q dn}.
\end{equation}
Substituting \( E_n \) from Eq.~\ref{eq:En} and changing variables to \( z \), we obtain:

\begin{equation} \label{int_01}
    U_q \simeq \frac{A^2}{4B}-\frac{1}{\beta(1-q)}\frac{\int_0^{z_{max}} z^2 \left[R^2+z^2\right]^{\frac{q}{1-q}}dz}{\int_0^{z_{max}} \left[R^2+z^2\right]^{\frac{q}{1-q}}dz},
\end{equation} where the $z$, $R$, $A$, and $B$ have the same meaning as used in the derivation of partition function in the moderate to low temperature limit.
Thus, Eq.~\ref{int_01} can be further written as:
\begin{equation}
    U_q \simeq \frac{A^2}{4B}-\frac{1}{\beta(1-q)}\frac{\int_0^{\frac \pi 2} (R \tan{\theta})^2 (R\sec{\theta})^{\frac{2q}{1-q}}R \sec^2{\theta} d\theta}{\int_0^{\frac \pi2} (R\sec{\theta})^{\frac{2q}{1-q}}R \sec^2{\theta} d\theta}
\end{equation}

\begin{equation}
U_q \simeq \frac{A^2}{4B} + \frac{R^2}{(q - 3)\beta} \simeq  \frac{A^2}{4B} - \frac{R^2}{(3-q)\beta}.
\end{equation}
This reveals the linear inverse temperature scaling typical of vibrational systems and the \( q \)-dependent shift due to nonextensivity.

For moderate temperature, due to the Tsallis formalism, we get the correction term $\frac{R^2}{(q-3)\beta}$. Hence, the modified internal energy has the following generalized form:
\begin{equation}\label{analytics_avg_eng}
U_q=\mathrm{const.}+\frac{g(q)}{\beta},
\end{equation}
where $g(q)=\frac{R^2}{(q-3)}.$ This correction term shows that at intermediate temperature, the internal energy is suppressed by an amount $\frac{R^2}{(3-q)\beta}$, and the suppression is more significant for smaller $q.$
\subsection{Deformed entropy \texorpdfstring{$S_q$}{Sq}}
For the sake of completeness, we briefly discuss the modified entropy (or Tsallis entropy) using the definition of Tsallis entropy in the continuous limit, and we write:
\begin{equation}
S_q = k_B \frac{1 - \int_0^{n_{\mathrm{max}}} \left[\frac{\exp_q(-\beta E_n)}{Z_q}\right]^q dn}{q - 1}.
\end{equation}

From earlier expressions for the integrals, this reduces to:
\begin{equation}
S_q = \frac{k_B}{q - 1} \left[
1 -
\frac{ R^{\frac{(1+q)}{(1-q)}} \Gamma\left(\frac{1}{2}\right)\Gamma\left(-\frac{1 + q}{2(1 - q)}\right) }
{ 2 \cdot Z_q^q \sqrt{(1 - q)\beta B} \cdot  \Gamma\left(-\frac{q}{1 - q}\right)}
\right].
\end{equation}
This expression encodes the nonextensive corrections in the thermodynamic measure of disorder of the system.
\section{Effect of Anharmonicity and Nonextensivity on Internal Energy}
 \label{sec:anharmonicity}

\begin{figure}[!htbp]
    \centering
    \includegraphics[width=1\textwidth]{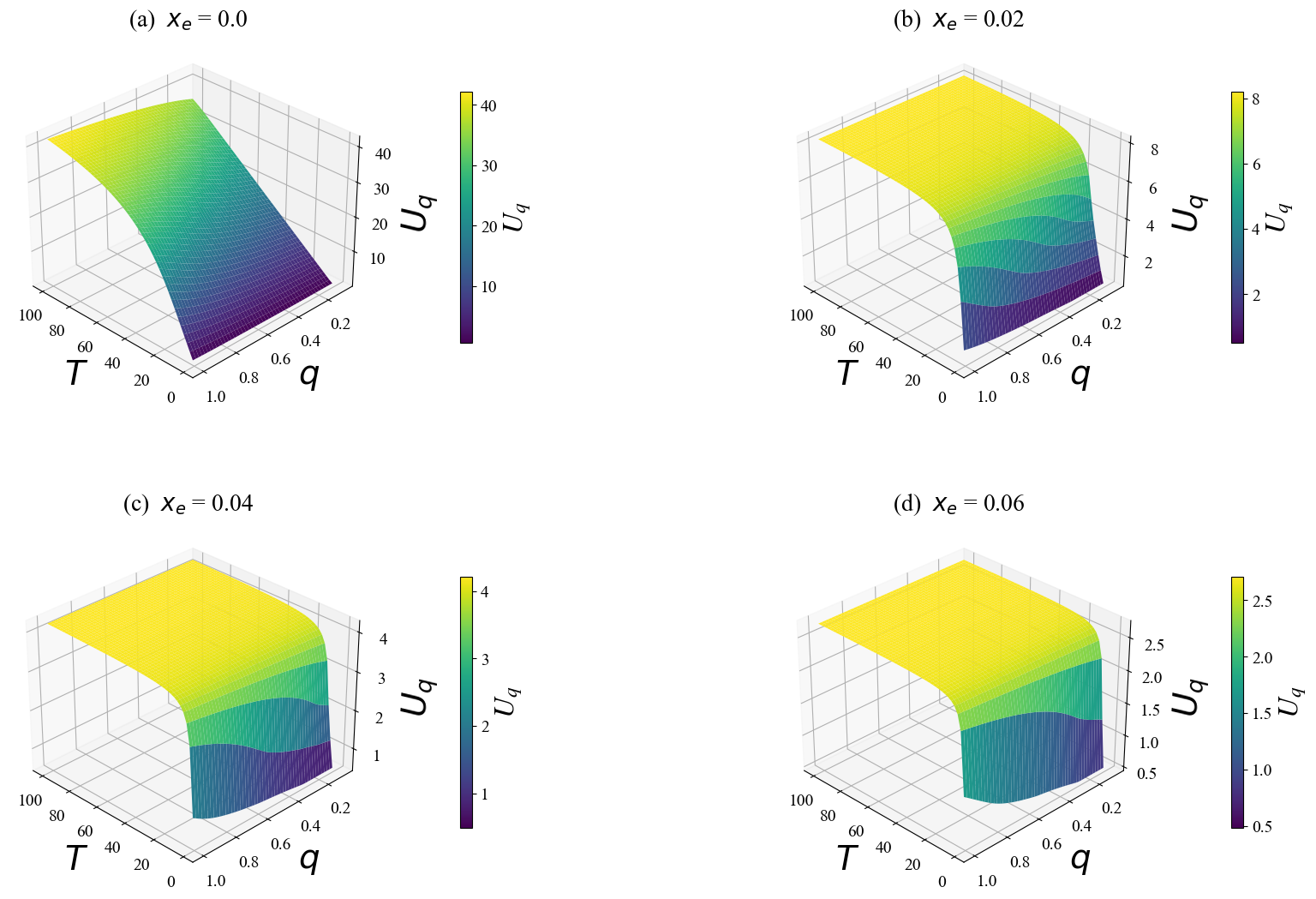} 
    \caption{Surface plots of internal energy \( U_q(T, q) \) for the Morse oscillator at increasing anharmonicity values: \( x_e = 0.00, 0.02, 0.04, 0.06 \) arranged from left to right and top to bottom. As \( x_e \) increases, the number of bound states decreases, leading to earlier saturation of internal energy. For \( q < 1 \), nonextensivity further suppresses contributions from high-energy levels due to the compact support of the \( q \)-exponential. The combined effect results in significant thermodynamic suppression, especially at intermediate to high temperatures. All panels $n_{\mathrm{max}}$ is adjusted as a function of \( x_e \). $k_BT$ and $U_q$ are in eV.}

    \label{fig:Uq_surface_xe}
\end{figure}

To understand how the combination of microscopic anharmonicity and statistical nonextensivity jointly affects thermal behavior, we calculate the internal energy \( U_q(T, q) \) of the Morse oscillator as a function of temperature \( T \), the Tsallis parameter \( q \), and the anharmonicity constant \( x_e \). Fig.~\ref{fig:Uq_surface_xe} presents a set of 3D surface plots illustrating the variation of the internal energy across the \((T, q)\) plane for different intensities of anharmonicity \( x_e \).

In the harmonic limit \( (x_e = 0) \), the oscillator spectrum is nearly equidistant and unbounded. As shown in the upper left panel of Fig.~\ref{fig:Uq_surface_xe}, the internal energy increases monotonically with temperature. The nonextensive effects, seen at lower values of \( q \), appear primarily as a suppression of high-temperature energy growth due to the reduced weight in higher-energy states.

With an increase in the anharmonicity parameter \( x_e \), the nonlinearity in the behavior of energy levels increases, and the number of bound states decreases. From Eq.~\ref{eq:En}, the maximum number of bound states can be written as:
\[
n_{\mathrm{max}} \sim \frac{1}{2x_e} - \frac{1}{2},
\]
which shows that a larger \( x_e \) leads to fewer energetically accessible levels. This spectral truncation results in a lower maximum internal energy and a slower thermal rise, especially pronounced for lower \( q \), where the Tsallis distribution further suppresses the population of higher energy states.

At strong anharmonicity \( (x_e = 0.06) \), shown in the bottom right panel, the internal energy exhibits early saturation with minimal temperature dependence, particularly in the subextensive regime \( q < 1 \). This saturation arises due to the combined effects of (i) a finite bound spectrum due to strong anharmonicity and (ii) compact support of the \( q \) exponential distribution, which statistically excludes high-energy levels.

Thus, the internal energy reflects a thermodynamic suppression driven by both microscopic structure and nonextensive statistical weighting. These results indicate the sensitivity of vibrational thermodynamics (thermodynamics of a diatomic molecule) to the interplay between energy spectrum topology and entropy deformation, especially in molecular systems with finite dissociation energies.

\section{Numerical Evaluation and Results} \label{sec4}

In this section, we numerically evaluate the generalized thermodynamic quantities for the Morse oscillator under Tsallis statistics in the subextensive regime \( q < 1 \). All calculations are performed using a finite number of bound states, \( n_{\mathrm{max}} = 49 \), and are compared with Boltzmann–Gibbs (BG) results for reference. For our numerical calculations, we take $\hbar \omega_0 =1~\mathrm{eV}$, and $D_e=25~\mathrm{eV}$ and $x_e=0.01.$

\subsection{Generalized partition function and probability distribution}

\begin{figure}[htbp]
    \centering
    \begin{subfigure}[t]{0.45\linewidth}
        \includegraphics[width=\linewidth]{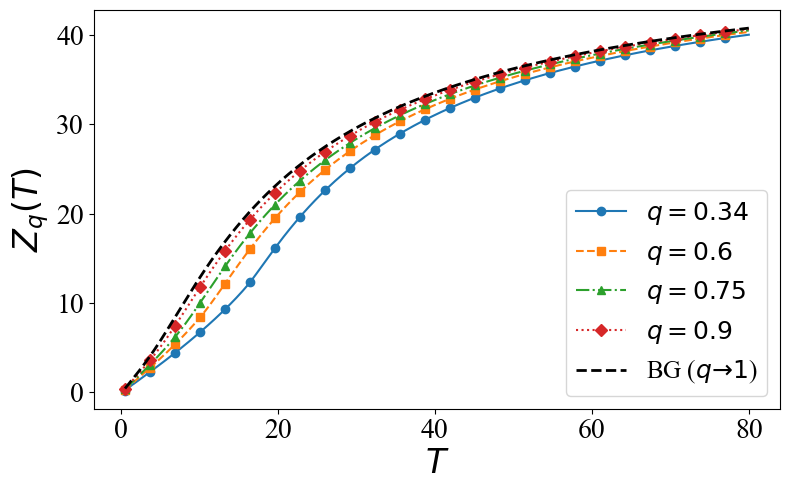}
        \caption{\( Z_q(T) \) vs. \( T \) for \( q < 1 \)}
    \end{subfigure}
    \hspace{0.05\linewidth}
    \begin{subfigure}[t]{0.45\linewidth}
        \includegraphics[width=\linewidth]{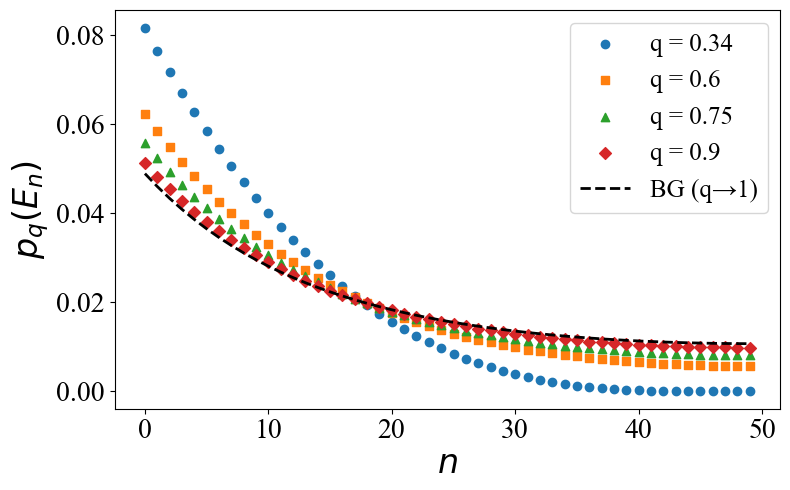}
        \caption{\( p_q(E_n) \) vs. \( n \) at \( k_BT = 15~\mathrm{eV} \).}
    \end{subfigure}
    \caption{(a) Generalized partition function \( Z_q(T) \) for different values of \( q < 1 \) as a function of temperature ($T$). (b) Probability distribution \( p_q(E_n) \) over energy levels \( n \) at fixed temperature. The parameters are as follows: $\hbar \omega_0 =1~\mathrm{eV}$, $D_e=25~\mathrm{eV}$ and $x_e=0.01,$ with $n_{\mathrm{max}}=49$. The plots illustrate that smaller q values narrow the distribution and modify the $Z_q(T)$ growth.}
    \label{fig:fig1}
\end{figure}

Fig.~\ref{fig:fig1}(a) shows the temperature dependence of the generalized partition function \( Z_q(T) \) for various values of \( q < 1 \). As expected, in the limit \( q \to 1 \), the standard BG result is recovered. For smaller \( q \), the partition function is suppressed at low (intermediate) temperatures, reflecting the compact support of the \( q \)-exponential, which limits the contribution of the high-energy states. 

Fig.~\ref{fig:fig1}(b) plots the corresponding probability distribution \( p_q(E_n) \) as a function of the quantum number \( n \) at a fixed thermal energy \( k_BT = 15 ~\mathrm{eV} \), which is in the proposed intermediate temperature scale. The plot shows that the probability of occupation of the high-energy states is lower. The lower value of q corresponds to the strongest sub-extensive behavior, and it largely suspends the contribution of the high-energy states. \par The results are in good agreement with the analytical prediction of the partition function in Eq.~\ref{partition_analytics} and the finding that the Tsallis statistics lower the upper bound (or cut-off state in Eq.~\ref{analytics_neff}). And $p_q(E_n)$ for high q ($q=0.9$) matches exactly with the BG statistics.
\subsection{Effective number of states and internal energy ratio}
\begin{figure}[t]
    \centering
    \begin{subfigure}[t]{0.45\linewidth}
        \includegraphics[width=\linewidth]{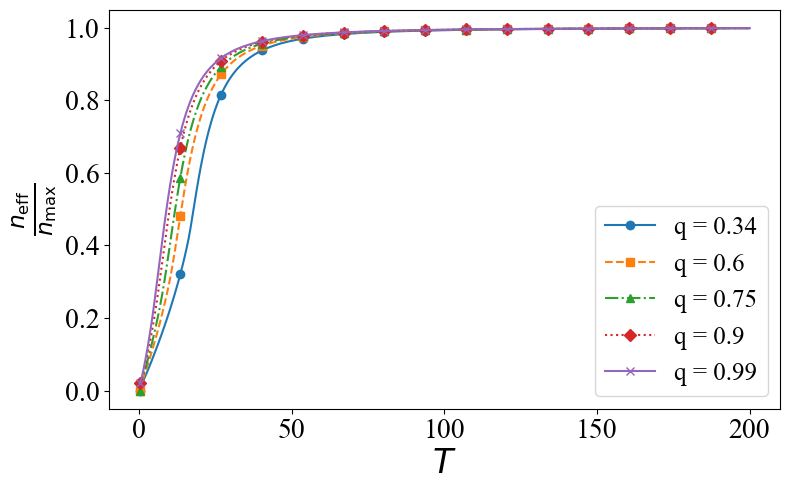}
        \caption{\( \frac{n_{\mathrm{eff}}} {n_{\mathrm{max}}} \) vs. \( T \)}
    \end{subfigure}
    \hspace{0.05\linewidth}
    \begin{subfigure}[t]{0.45\linewidth}
        \includegraphics[width=\linewidth]{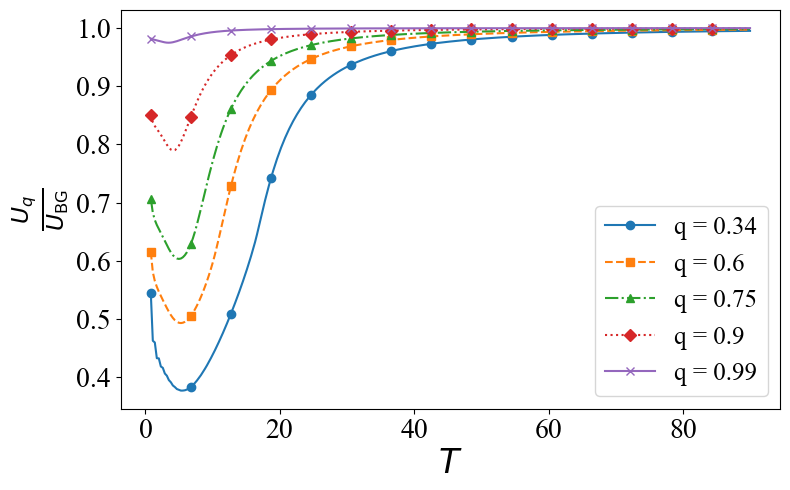}
        \caption{\( \frac{U_q} {U_{\mathrm{BG}}} \) vs. \( T \)}
    \end{subfigure}
    \caption{(a) Temperature dependence of the fraction of effectively contributing states \( \frac{n_{\mathrm{eff}}} {n_{\mathrm{max}}} \). (b) Ratio of internal energy in Tsallis and BG frameworks \( \frac{U_q} {U_{\mathrm{BG}}} \), showing a dip due to nonextensive suppression of higher-energy states. $k_B T$ is in eV. The parameters are as follows: $\hbar \omega_0 =1~\mathrm{eV}$, $D_e=25~\mathrm{eV}$ and $x_e=0.01,$ with $n_{\mathrm{max}}=49$.}
    \label{fig:fig2}
\end{figure}

We now analyze two key indicators of nonextensive thermodynamic behavior: the fraction of effectively contributing states \( n_{\mathrm{eff}} \), and the normalized internal energy \( \frac{U_q } {U_{\mathrm{BG}}} \). The results are shown in Fig.~\ref{fig:fig2}.

Fig.~\ref{fig:fig2}(a) displays the normalized effective number of contributing states, \( \frac{n_{\mathrm{eff}}}{n_{\mathrm{max}}} \) as a function of temperature for different values of the Tsallis parameter \( q \). At low temperatures, only a limited number of bound states make a significant contribution to thermodynamics. Interestingly, the number of states that make an effective contribution increases as \( q \) increases, indicating that nonextensive effects with \( q < 1 \) promote the possibility for low-lying energy levels to participate in thermal properties. This behavior arises from the shape of the \( q \)-exponential distribution: for \( q < 1 \), it becomes flatter over the low-energy spectrum, enhancing the occupation of these states before truncating at finite energy due to compact energy support.

\begin{figure}[!ht]
    \centering
    \includegraphics[width=0.5\linewidth]{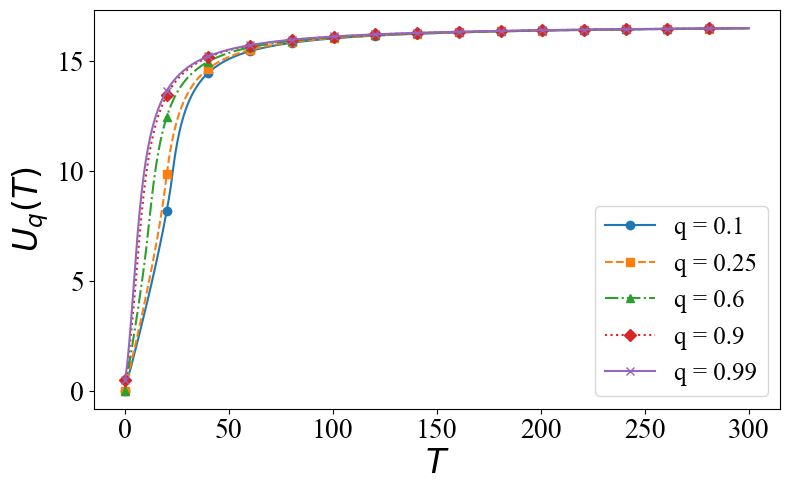}
    \caption{Plot for generalized internal energy $U_q(T)$ as a function of temperature ($T$) for different $q$ values. $U_q$ and $k_BT$ are in eV. The parameters are as follows: $\hbar \omega_0 =1~\mathrm{eV}$, $D_e=25~\mathrm{eV}$ and $x_e=0.01,$ with $n_{\mathrm{max}}=49$. The difference in $U_q(T)$ occurs in the intermediate-temperature region and vanishes at high temperature, and collapses with the BG prediction.}
    \label{fig:int_01}
\end{figure}

Fig.~\ref{fig:fig2}(b) shows the ratio \( \frac{U_q }{ U_{\mathrm{BG}}} \) as a function of temperature. This quantity exhibits a distinct drop at intermediate temperatures for all \( q < 1 \). The origin of the dip is as follows: at low temperatures, both the Tsallis and BG distributions primarily populate the lowest-energy states, yielding similar internal energies. At intermediate temperatures, the BG framework allows higher-energy states to be occupied due to the exponential tail, whereas the Tsallis distribution cuts off contributions beyond a finite energy. This results in a slower growth of \( U_q \) relative to \( U_{\mathrm{BG}} \) at a given temperature, creating the observed drop (as predicted from the analytical expression in Eq.~\ref{analytics_avg_eng}). At high temperatures, the effective energy cutoff becomes large, which makes the whole spectrum accessible to both distributions, and the ratio approaches unity:
\[
\lim_{T \to \infty} \frac{U_q}{U_{\mathrm{BG}}} = 1.
\]
This dip in the \( \frac{U_q}{U_{\mathrm{BG}}} \) plot suggests the possibility of seeing some anomaly in the specific heat at the intermediate temperature. Particularly for Fig.~\ref{fig:fig2}(b), we started from a very small non-zero $T$ in order to eliminate the divergence at $T=0$. We also plotted the average internal energy ($U_q$) as a function of temperature at different q values, which is presented in Fig.~\ref{fig:int_01}. The plot shows a similar behavior to the plot of $\frac{n_{\mathrm{eff}}}{n_{\mathrm{max}}}$ as a function of T.
\subsection{Tsallis entropy and specific heat plots}
\begin{figure}[t]
    \centering
    \begin{subfigure}[t]{0.45\linewidth}
        \includegraphics[width=\linewidth]{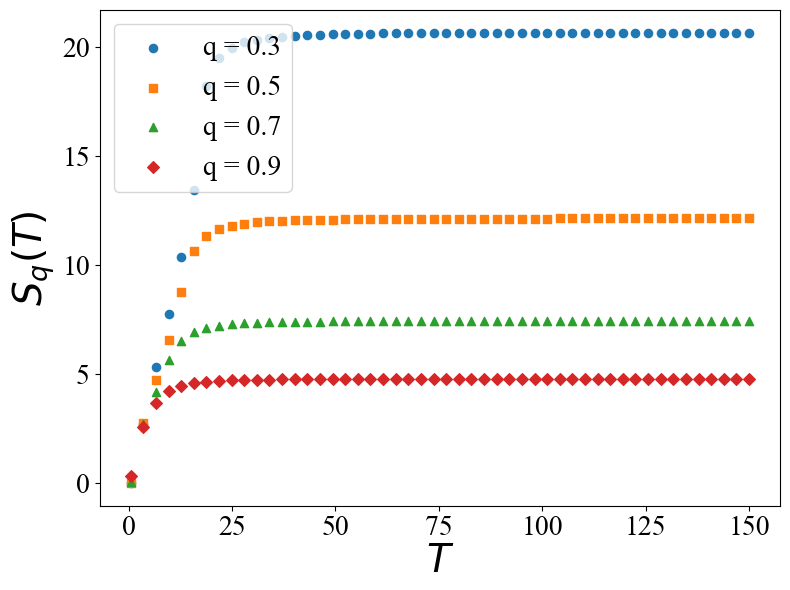}
        \caption{\( S_q(T) \) vs. \( T \) for different \( q \) values.}
    \end{subfigure}
    \hspace{0.05\linewidth}
    \begin{subfigure}[t]{0.45\linewidth}
        \includegraphics[width=\linewidth]{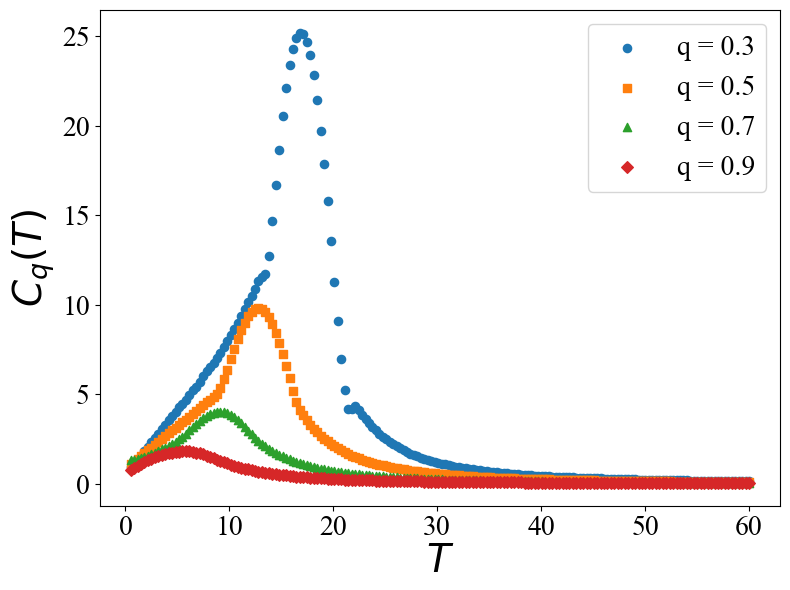}
        \caption{\( C_q(T) \) vs. \( T \) for different \( q  \) values.}
    \end{subfigure}
    \caption{(a) Tsallis entropy \( S_q(T) \), (b) Specific heat $C_q(T)$ for different values of \( q < 1 \) has been plotted as a function of temperature (T). The parameters are as follows: $\hbar \omega_0 =1~\mathrm{eV}$, $D_e=25~\mathrm{eV}$ and $x_e=0.01,$ with $n_{\mathrm{max}}=49$. Entropy shows $q$ dependent saturation, and Specific heat shows $q$ dependent peak position as a signature of nonextensivity. $k_B T$ is in eV.}
    \label{fig:figapp}
\end{figure}
To have a complete and concrete discussion, we also present the graphs for generalized entropy ($S_q(T)$) and generalized specific heat ($C_q(T)$) in Fig. \ref{fig:figapp}(a), and \ref{fig:figapp}(b), respectively. Here, the $C_q(T)$ plots at different $q$-values show peaks at different temperatures, which is a signature of Schottky-type anomalies in systems with finite bound states (discrete energy levels). This situation arises due to the interplay between nonextensivity and limited accessible states (or phase-space). Similar behavior has been observed in systems with discrete energy levels\cite{Lee2015}. \par The generalized Tsallis entropy $S_q(T)$ vs. T plots show a saturation at high temperatures. We also see that for different q values, the saturation value at high temperature is different; at small q values, the low-energy states contribute more to the thermal property. Similar entropy and specific heat behavior under Tsallis statistics have been observed for the system of multiple quantum harmonic oscillators \cite{Ishihara2024}. This generalized Tsallis entropy is the measure of the deformation of generalized theories (theory corresponds to $q<1$) from the equilibrium BG formalism. And, as we know, the lower the value of q, the greater the deviation that the theory shows. For this reason, the entropy corresponding to $q=0.3$ is more than the entropy of $q=0.9.$
\begin{figure}[t]
    \centering
    \begin{subfigure}[t]{0.45\linewidth}
        \includegraphics[width=\linewidth]{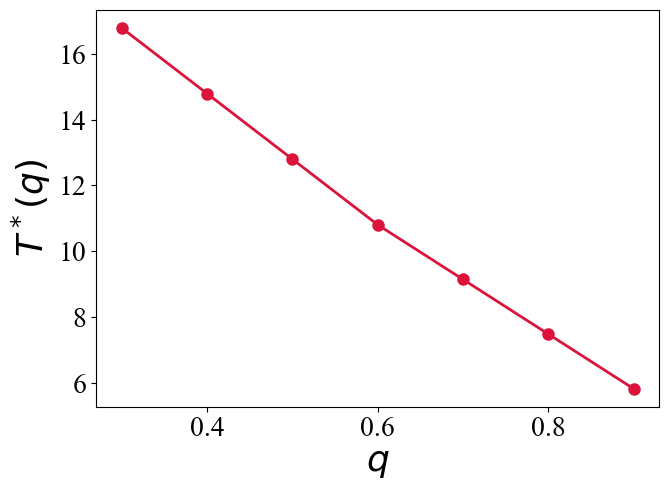}
        \caption{\( T^*(q) \) vs. \( q \)}
    \end{subfigure}
    \hspace{0.05\linewidth}
    \begin{subfigure}[t]{0.45\linewidth}
        \includegraphics[width=\linewidth]{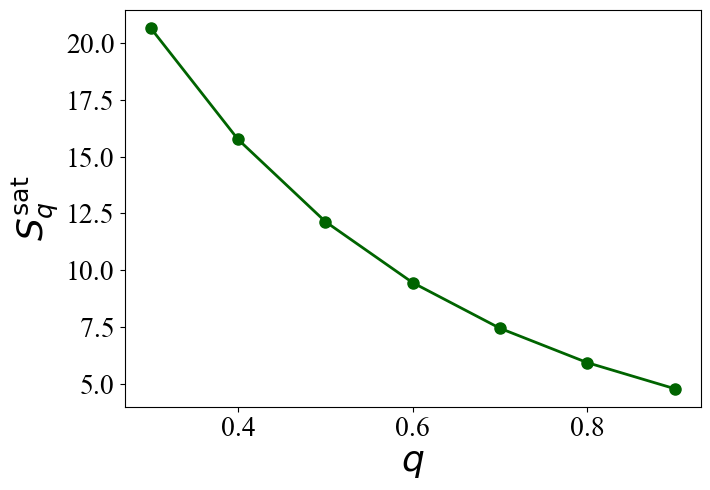}
        \caption{\( S_q^{sat} \) vs. \( q \) }
    \end{subfigure}
    \caption{(a) Peak position of specific heat \( T^*(q) \), (b) Saturation entropy $S^{sat}_q$ has been plotted as a function of $q.$ The monotonic decay of both $T^*(q)$ and $S^{sat}_q$ indicates the subextensive nature of the statistics. {The parameters are as follows: $\hbar \omega_0 =1~\mathrm{eV}$, $D_e=25~\mathrm{eV}$ and $x_e=0.01,$ with $n_{\mathrm{max}}=49$.}}
    \label{figll}
\end{figure}

\par
One can define $T^*(q)$, the temperature at which the specific heat $C_q(T)$ reaches its maximum. Its value is different for different values of $q$. Fig.~\ref{figll}(a) shows that the Schottky-like anomaly shifts toward higher temperatures with decreasing $q$. This behavior can be explained by the behavior of the internal energy. For smaller q values at low temperatures, only a few states will contribute to the internal energy. After that, with an increase in temperature, the number of contributing states will increase, and at some temperature, the number will hit its maximum limit because of the bound state property. After that, with further increase in temperature, there will be no further increase, and it will remain constant. The temperature at which it hits the maximum limit is the temperature at which the nonextensive theory becomes comparable with the BG statistics. The peak actually signifies the transition from a particular q-statistics to the BG statistics. The smaller $q$ values denote strong nonextensivity. So, the smaller the value of $q$, the higher the thermal energy needed to recover the BG statistics. This behavior matches our theoretical prediction in Eq.~\ref{T_scal}.

 $S_q^{\mathrm{sat}}$ is defined as the saturation value of the entropy in the high-temperature limit and shows a monotonic decrease with $q$. Fig.~\ref{figll}(b) reflects the fact that as $q$ decreases, the effective number of thermally accessible microstates also becomes much less than the number of states expected in BG statistics, leading to a greater deviation from the BG statistics. As a result, $q \ll 1$, the deformation entropy is notably higher even at high $T$. In the BG limit, the deformation entropy reaches its minimum value. Together, the $T^*(q)$ and $S_q^{\mathrm{sat}}(q)$ plots provide complementary insights into how nonextensivity modulates both the thermal activation scale and the asymptotic deformation entropy content in quantum systems with bounded spectra.
 \par Now the obvious question that arises is that, at high temperature, the results of the $q$ statistics should match with the BG statistics, then why, for a large temperature, the entropy is higher for smaller $q$ and lower for larger $q$. To understand this, we have to understand the form of the deformation entropy in detail. By definition, the form of the Tsallis entropy is given as:
 \begin{equation}\label{entropy} S_q=k\frac{1-\sum_{i=1}^{W}p^q_i}{q-1}.\end{equation} At high temperature, all states are equally probable for the system, so $p_i=\frac{1}{W}$. Now, if we substitute this in Eq.~\ref{entropy}, then we get the following:
 \[S_q =k\frac{(1-W\cdot(\frac{1}{W})^q)}{(q-1)}=k\frac{(1-W^{(1-q)})}{(q-1)}\]
This shows that for the limit $q\ne 1$, even at the high temperature limit, the entropy is dependent on $q$.
\section{Observable effects of the q-deformation \label{new_sec}}
In summary, this study gives a rigorous q-deformed thermodynamic framework; however, it is equally important to interpret the mathematical constructs in terms of the physical behavior of real observables. There are many known quantifiers in information theory where one can indeed see the effect of any changes of parameter values (q in our case) on the thermodynamic properties of the system, and they are experimentally realizable quantities. Fisher information is one of the quantifiers that is known to be often associated with the sensitivity in the probability distribution of a system to the changes in parameters \cite{frieden2004science}, which may reflect the sharpness of vibrational transitions or the localization of energy states that are potentially accessible via high-resolution spectroscopy. On the other hand, entropy production, which is known as the quantifier of irreversibility and deviation from equilibrium \cite{VANDENBROECK20156}, can be realized through relaxation dynamics or heat dissipation patterns in solid-state systems. At last, thermodynamic curvature, which arises from information geometry, and affects interaction strength and stability \cite{Janyszek_1990}, can possibly be detected in the anomalous nature of heat capacity or susceptibility near critical points. These suggest that the deformation parameter q not only modifies statistical behavior but also shows measurable effects on the system’s physical response. This study derives the q-probability distribution, Tsallis entropy, and generalised heat capacity that reflect similar aspects of Fisher information, entropy production, and curvature, respectively. \\

\section{Estimating specific heat of solids with the Morse oscillator potential}\label{secapp} Now, as we establish the thermodynamic framework of the q-deformed Morse oscillator, we will next tend to explore its relevance in solid-state contexts in the following sections. Specifically, we will explore how the deformation parameter q influences the vibrations of the lattice and the energy distributions in crystalline solids. \\
As we discussed earlier, in this part of the paper, we study one of the most fundamental problems of statistical physics, the specific heat of a solid. We used the discrete energy levels of the Morse oscillator for this study. The benefit of using this anharmonic oscillator system is that it can take into account thermal fluctuations at room temperature and high temperature limits. So, considering the Morse oscillator energy levels as phononic modes of vibration, we can get a generalized expression of specific heat in solids. We will go with the more useful Debye approach \cite{kittel2004solid}. To do that, first, we will try to linearize the potential. We have the following potential expression for the Morse oscillator system:
\[V(x) =D_e \left[1-e^{-a(x-x_0)}\right]^2.\]
Considering $x-x_0 =\xi$, we get the following expression:
\[V(\xi) =D_e \left[1-2e^{-a\xi} + e^{-2a\xi}\right].\]
We can expand the exponential function:
\begin{eqnarray}
V(\xi) & =& D_e \bigg[
    1 
    - 2\left(1 - a\xi + \frac{(a\xi)^2}{2!} - \frac{(a\xi)^3}{3!} + \frac{(a\xi)^4}{4!}\right) \nonumber \\
&& \quad + 1 - 2a\xi + \frac{(2a\xi)^2}{2!} - \frac{(2a\xi)^3}{3!} + \frac{(2a\xi)^4}{4!}
\bigg]
\end{eqnarray}

which can be further reduced as:
\begin{equation} \label{potential}
    V(\xi)=D_e\left[(a\xi)^2 -(a\xi)^3+\frac{7}{12} (a\xi)^4\right].
\end{equation}
From Eq.~\ref{potential}, we can calculate the effective force due to the stretch $\xi$:\[F=-\frac{\partial V}{\partial \xi}= -D_e(2a^2 \xi-3a^3\xi^2 +\frac{7}{3}a^4\xi^3),\] where $F$ can be expressed as: $F=m\frac{\partial^2 \xi}{\partial t^2},$ and $m$ represents the mass.
Let, $\xi(t)=A_0\cos{\omega t},$ where $A_0$ represents the amplitude. Substituting this into the above equation, we get:
\[-m\omega^2 A_0\cos{\omega t}=-D_e(2a^2A_0 \cos{\omega t}-3a^3A_0^2\cos^2{\omega t}+\frac{7}{3}a^4 A_0^3\cos^3{\omega t})\]
In the above equation, substituting $\cos^2{\omega t}=\frac12(1+\cos{}2\omega t)$ and $\cos^3{\omega t}=\frac14(3\cos{\omega t}+\cos{3\omega t})$, and by equating the coefficients of $\cos{\omega t}$ we get:
\[\omega^2 = \frac{D_e}{m}(2a^2+\frac74 a^4 A_0^2),\] where the natural frequency of the morse oscillator is $\omega_0=\sqrt{\frac{2D_e a^2}{m}},$ and let $\beta=D_e a^4.$ Thus, we get the following expression:
\begin{equation}
    \omega=\omega_0 \left[1+\frac{7\beta A_0^2}{8m\omega^2_0}\right]
\end{equation}
Now, from the equipartition principle, we can approximately write 
\[ \frac12 m \omega^2_0 A_0^2 = \frac12 k_B T\] From this we get $A_0^2=\frac{k_B T}{m \omega^2_0}.$ Thus, the effective frequency takes the following form:
\[\omega =\omega_0 \left[1+\frac78 \frac{D_e a^4 (k_B T)}{(m\omega^2_0)^2} \right] = \omega_0 \left[1+ \frac{7(k_B T)}{32D_e} \right].\] This can be written in a more general form as follows (since the natural frequency $\omega_0$ is only a function of $k$, where $k$ is the wave vector):
\begin{equation}\label{gen_dis}
    \omega(k,T)=\omega_0(k)\left[1+\alpha(T)\right]
\end{equation} where $\alpha(T)=\frac{7 (k_B T)}{32 (D_e)}.$ In our analysis we are mostly interested in the $\hbar \omega_0 <k_B T < D_e ~(\mathrm{intermediate ~ temperature})$ and $k_B T \simeq D_e ~(\mathrm{high ~ temperature}),$ that make $\alpha(T) <1.$ Now from the standard dispersion relation we know that $\omega_0 \propto k ~(\mathrm{for~ small~ k}).$ This reduces the general dispersion relation in Eq.~\ref{gen_dis} as follows:
\begin{equation}
    \omega \propto k (1+\alpha(T))
\end{equation}
From this general expression of the dispersion relation, one can understand that now the Debye frequency ($\omega_D$) is going to be a temperature-dependent quantity, and in the conventional density of state expression, we just need to replace $\omega_D$ with $\omega_D(T)$. Keeping that in mind, we can write the number of states within the frequency range $\omega$ to $\omega +d\omega$ ($k$ to $k+dk$) as follows:
\[g(\omega) d\omega=\frac{9N \omega^2}{\omega_D^3(T)} d\omega,\] where $N$ is the number of atoms in the solid. Thus, the internal energy can be expressed as:
\begin{equation}
    U(T)=\int_{0}^{\omega_D(T)} \hbar \omega \langle n(\omega) \rangle g(\omega) d\omega,
\end{equation}
where $\langle n(\omega) \rangle$ is the average excitation quantum number of the oscillator, which can be expressed as the following q-generalized form:
\begin{equation}
   \langle n_q(\omega) \rangle =\frac{\sum_{n} n\exp_q{(-n\hbar \omega \beta)}}{\sum_{n} \exp_q{(-n\hbar \omega \beta)}}
\end{equation}
where \[\exp_q{(-n\hbar \omega \beta)}=\left[1-(1-q)n \hbar \omega \beta\right]^{\frac{1}{(1-q)}}.\] In the following section, we will derive the detailed expression of the specific heat of solids using the Debye approach.
\subsection{Specific heat using BG statistics} In this section, we will calculate the specific heat expression from the BG statistics.
For the limit $q \to 1$, the generalized average excitation quantum number takes the following form:~$\langle n(\omega) \rangle =\frac{1}{(\exp{(\beta \hbar \omega)}-1)}.$ Using this expression of $\langle n(\omega)\rangle$, we calculate the internal energy expression as follows:
\[  U(T)=\int_{0}^{\omega_D(T)} \hbar \omega  \frac{1}{(\exp{(\beta \hbar \omega)}-1)} \frac{9N \omega^2}{\omega_D^3(T)}d\omega\]
The variable transformation:~$x=\beta \hbar \omega$, can simplify the expression in the following form:
\begin{equation}  U(T)=9Nk_B T{\bigg(\frac{T}{\theta_D(T)}\bigg)}^3\int_{0}^{\bigg(\frac{\theta_D(T)}{T}\bigg)}  \frac{x^3}{(\exp{(x)}-1)} dx\label{int_eng_0},\end{equation} where $\theta_D(T)=\frac{\hbar \omega_D(T)}{k_B}.$ Solving the integration exactly is difficult, so we separately consider two temperature regions to obtain expressions.
\subsubsection{Low-temperature limit:} First, we will discuss the low-temperature case. It should be noted that we are not working in the extremely low temperature limit for which $k_B T \ll \hbar \omega_0.$ That is because at that temperature either $n=0$ or $n=0,~1$ states will participate in the thermal property of the system, and that will not be quite helpful for seeing the combined effect of the bound spectra and the nonextensivity. Thus, here by low temperature we refer to the limit $\hbar \omega_0 < k_B T << D_e$. Now, at low temperature the upper limit of integration can be approximated to $\infty$ (as $\theta_D(T) >>T$), and thus the integration takes a finite value: \[\int_{0}^{\infty} \frac{x^3}{\exp{(x)}-1} dx=\frac{\pi^4}{15}.\] Substituting this integration value in Eq.~\ref{int_eng_0} we get the following expression:
\begin{equation}
    U(T)=9Nk_BT\bigg(\frac{T}{\theta_D(1+\alpha(T))}\bigg)^3 \frac{\pi^4}{15}.
\end{equation}
From this internal energy expression, we get the following specific heat expression:
\begin{equation}\label{C_BG}
C_v=\frac{9Nk_B \pi^4}{15\theta_D^3}(4T^3-15cT^4),\end{equation} where c is the coefficient of the correction term which arises due to the anharmonicity of the potential of the Morse oscillator, and the value of c is $\frac{7}{32}(\frac{k_B }{D_e}).$ At extremely low temperature ($T \to 0$) $C_v \to 0$, consistent with the thermodynamic laws.

\subsubsection{High-temperature limit:}

In the high-temperature limit ($k_B T \simeq D_e$), the expression of the internal energy in Eq.~\ref{int_eng_0} can be reduced as:
\[U(T) =9Nk_B T \bigg(\frac{T}{\theta_D (T)}\bigg)^3 \int_{0}^{\bigg(\frac{\theta_D}{T}\bigg)} x^2 dx,\]
after performing the integration, which can be further reduced to:
\begin{equation}
    U(T) =3Nk_B T,
\end{equation}
This gives the $C_v$ expression at high temperature as $C_v = 3Nk_B.$ This shows that the high-temperature specific heat remains unaffected (follows Dulong-petit law) by introducing a nonlinear term in the potential of the oscillator; however, the $C_v$ at intermediate temperature changes significantly, which appears as the correction term in the $C_v$ expression in the previous section.
\subsection{Specific heat using Tsallis statistics}
Now, we will use Tsallis statistics to calculate the specific heat expression of the solids. To do that, first we need to find the generalized expression for the average excitation quantum number of the oscillator in the q-generalized statistics: \begin{equation}\langle n_q(\omega) \rangle =\frac{\sum_{n=0}^{\infty} n\bigg[1- (1-q)n\hbar \omega \beta\bigg]^{\frac{1}{(1-q)}}}{\sum_{n=0}^{\infty} \bigg[1- (1-q)n\hbar \omega \beta\bigg]^{\frac{1}{(1-q)}}}. \label{avg_n_q}\end{equation}
Let $(1-q)\hbar \omega \beta=x$, then $\bigg[1- (1-q)n\hbar \omega \beta\bigg]^{\frac{1}{(1-q)}}=(1-nx)^{\frac{1}{(1-q)}}.$ Substituting this in Eq.~\ref{avg_n_q} we get as follows:

\[n_q(\omega) \rangle =\frac{\sum_{n=0}^{\infty} n(1-nx)^{\frac{1}{(1-q)}}}{\sum_{n=0}^{\infty}(1-nx)^{\frac{1}{(1-q)}}}.\]

Now, from the structure of the probability distribution in the Tsallis statistics, we can understand that the formalism will be valid only if $nx<1$ (as if $nx<1$ then $p_q(x)=(1-nx)^{\frac{1}{(1-q)}} > 0$). As we should have $nx<1$ so we can write $(1-nx)^{\frac{1}{(1-q)}}= \bigg((1-x)^{\frac{1}{(1-q)}}\bigg)^n.$ Thus, Eq.~\ref{avg_n_q} reduces to \[\langle n_q(\omega) \rangle =\frac{\sum_{n=0}^{\infty} n (\exp_q(-x))^n}{\sum_{n=0}^{\infty}(\exp_q(-x))^n}= \frac{1}{(\exp_q(\beta \hbar \omega)-1)}.\]
Thus, the internal energy in Eq.~\ref{int_eng_0} takes the following form:
\begin{equation} \label{avg_eng_0q} U(T)=\int_{0}^{\omega_D(T)} \hbar \omega  \frac{1}{(\exp_q(\beta \hbar \omega)-1)} \frac{9N \omega^2}{\omega_D^3(T)}d\omega\end{equation}
 Like in the BG case, here we will also derive $C_v$ at both the temperature limits using Tsallis statistics.
 \subsubsection{Low temperature limit:}
 
 Now, since we are mainly interested in the temperature (T), $\hbar \omega <k_B T<D_e$, so $\exp_q(-\beta \hbar \omega)$ can be approximated as follows: \[\exp_q(-\beta \hbar \omega)=\bigg[1-(1-q)\beta \hbar \omega\bigg]^{\frac{1}{(1-q)}}=\exp{\bigg[\ln{\bigg[1-(1-q)\beta \hbar \omega\bigg]^{\frac{1}{(1-q)}}}}\bigg],\]
where
 \[\exp{\bigg[\ln{\bigg[1-(1-q)\beta \hbar \omega\bigg]^{\frac{1}{(1-q)}}}}\bigg]=\exp{\bigg[\frac{\ln{\bigg[1-(1-q)\beta \hbar \omega\bigg]}}{(1-q)}\bigg]}.\] As we know $(1-q)\beta \hbar \omega <1$, we can have the following series expansion \[\ln\bigg[1-(1-q)\beta \hbar \omega\bigg]   \simeq -(1-q)\beta \hbar \omega -\frac{((1-q)\beta \hbar \omega)^2}{2}.\] Thus, \[\exp_q(-\beta \hbar \omega)=\exp{(-\beta \hbar \omega)\bigg(1-\frac{(1-q)(\beta \hbar \omega)^2}{2}+ \frac{(1-q)^2(\beta \hbar \omega)^4}{8}\bigg)},\] and the internal energy in Eq.~\ref{avg_eng_0q} takes the following form: \begin{equation}  \label{int_avg_0_1q}U(T)=\int_{0}^{\omega_D(T)} \hbar \omega  \frac{\exp_q(-\beta \hbar \omega)}{(1-\exp_q(-\beta \hbar \omega))} \frac{9N \omega^2}{\omega_D^3(T)}d\omega.\end{equation} Substituting the value of $\exp_q(-\beta \hbar \omega)$ in Eq.~\ref{int_avg_0_1q} we get the following expression:
 
\[U(T) \simeq\int_{0}^{\infty} \hbar \omega  \frac{(1-\frac{(1-q)(\beta \hbar \omega)^2}{2}+\frac{(1-q)^2(\beta \hbar \omega)^4}{8})}{(\exp{(\beta \hbar \omega)}-1)} \frac{9N \omega^2}{\omega_D^3(T)}d\omega.\]
This leads to the following:

\begin{eqnarray}
    U(T) &=& 9N k_BT \bigg(\frac{T}{\theta_D (1+\alpha(T))}\bigg)^3\bigg[\int_{0}^{\infty}  \frac{x^3}{(\exp{(x)}-1)}dx \nonumber \\
&& \quad -\frac{(1-q)}{2}\int_{0}^{\infty}  \frac{x^5}{(\exp{(x)}-1)}dx  \nonumber \\
&& \quad+ \frac{(1-q)^2}{8}\int_{0}^{\infty}  \frac{x^7}{(\exp{(x)}-1)}dx\bigg].
\end{eqnarray}
Now, using the finite values of the integral: \[\int_0^\infty \frac{x^n}{e^x - 1} \, dx = \Gamma(n+1)\zeta(n+1),\](where $\Gamma$ is the gamma function and $\zeta$ is the  zeta function respectively) we get the following:
\begin{equation}\label{int_02q}U(T) =9N k_BT \bigg(\frac{T}{\theta_D (1+\alpha(T))}\bigg)^3\bigg[ \frac{\pi^4}{15}-\frac{(1-q)}{2} \frac{8\pi^6}{63}+\frac{(1-q)^2}{8}\frac{8\pi^8}{15}\bigg]\end{equation}
 We can calculate the specific heat expression from the internal energy expression in Eq.~\ref{int_02q}:
 \begin{equation} \label{C_TS}
     C_v=A_qT^3(4-15c T)
 \end{equation}
 where $A_q=\frac{9Nk_B }{\theta_D^3}\bigg[ \frac{\pi^4}{15}-\frac{4(1-q) \pi^6}{63}+\frac{(1-q)^2\pi^8}{15}\bigg]$. It is worth mentioning here that as we take the perturbative expansion, this expression will be valid for the immediate neighbor of $q=1$ (approximately for $0.9<q<1$). To obtain the specific heat expression at low temperature for $q<<1$, one needs to perform a more sophisticated calculation, as the perturbation theory will become inadequate for that limit. The expansion also shows that $q = 1$ gives the results of the BG statistics.
 \subsubsection{High-temperature limit:}
 At sufficiently large temperature, we can have the following expansion for the general q-exponential function:
 \[\exp_q(\beta \hbar \omega)=\bigg[1+(1-q)\beta \hbar \omega\bigg]^{\frac{1}{(1-q)}} \simeq 1+\beta \hbar \omega,\] where the higher-order beta terms have been neglected. Thus, the internal energy takes the form:
 \[U(T) =9Nk_B T \bigg(\frac{T}{\theta_D (T)}\bigg)^3 \int_{0}^{\bigg(\frac{\theta_D(T)}{T}\bigg)} x^2 dx.\] This gives the specific heat expression as follows:
 \begin{equation}U=3Nk_B T, \quad C_v = 3Nk_B.\end{equation}
 This means that a correction arising from the Tsallis statistics will not have any effect on the high-temperature value of $C_v$, and we get the same results as Dulong-Petit law.
\par Here it is to be noted for the limiting case where the coupling constant $x_e \to 0$ (as this limit gives harmonic approximation of morse oscillator), then the $C_v$ in Eq.~\ref{C_BG} will reduce to the expression derived by Debye as the constant $c \to 0$, and Eq.~\ref{C_TS} becomes q-deformed Debye case. Importantly, the temperature scaling of $C_v$ remains valid in the thermodynamic limit $N \to \mathrm{large}$.
\section{Conclusion and Discussion} \label{sec5}

In this work, we presented a detailed analysis of the thermodynamic properties of the Morse oscillator within the framework of Tsallis nonextensive statistics. We restrict our discussion to the subextensive regime \( q < 1 \). The Morse potential naturally features a finite number of bound energy levels due to its anharmonic structure, allowing us to impose an intrinsic cutoff without enforcing any artificial truncation of the spectrum. This makes the Morse oscillator a physically realistic and analytically tractable model for studying nonextensive effects in bounded quantum systems. 

We derived analytical approximations for the generalized partition function \( Z_q(T) \) at both low- and high-temperature limits using asymptotic and continuum techniques. These expressions consistently recover the Boltzmann–Gibbs (BG) results in the limit \( q \to 1 \), as expected. In addition to internal energy and entropy, we introduced and analyzed two important quantities: the effective number of contributing states \( n_{\mathrm{eff}} \), and the normalized internal energy \( \frac{U_q} {U_{\mathrm{BG}}} \), both of which serve as sensitive indicators of nonextensivity.

The numerical evaluations are in good agreement with our analytical findings and revealed distinguishing features associated with the Tsallis formalism. At low temperatures, the effective number of contributing states \( n_{\mathrm{eff}} \) is significantly reduced for smaller \( q \), reflecting the subextensivity of the \( q \)-exponential distribution. The internal energy ratio \( \frac{U_q} {U_{\mathrm{BG}}} \) shows a clear depletion at intermediate temperatures, which is an immediate consequence of the suppression of occupation of the higher energy state in the Tsallis framework. At high temperatures, all thermodynamic quantities converge smoothly to their BG counterparts, affirming thermodynamic consistency. We also used a Morse oscillator to study one of the most fundamental problems of condensed matter physics: the specific heat of solids. This approach incorporates anharmonicity into the solid-state thermodynamics.

This work demonstrates how the coexistence of confined spectra and nonextensivity can cause suppression effects in many physically important thermodynamic parameters.  Our analysis suggests the Morse oscillator is one of the minimal models that can be used to analytically and numerically represent the behavior of nonextensivity.  This work has potential implications in the fields of confined quantum systems, vibrational spectroscopy, and molecular thermodynamics.

The future scope of this work includes:
\begin{itemize}
    \item To extend the model to more complex molecules, one can include rotational motion and different coupling effects.
    \item One can explore dynamical observables in the nonextensive regime.
    \item Investigating quantum information-theoretic measures, such as Tsallis-based entanglement entropy or Fisher information;
    \item This framework can be applied to experimental systems, such as cold molecular gases, anharmonic traps, or astrophysical plasmas, where nonextensivity may naturally arise.
\end{itemize}
Theoretical predictions presented in this work could be tested in controlled experimental settings for molecular systems, spectroscopic measurements of vibrational energy levels in diatomic species such as H$_2$, N$_2$, or CO —particularly at intermediate temperatures where anharmonic effects are pronounced. These offer a direct route for comparison with our $q$-dependent thermodynamic quantities. Our extension of the Morse model to specific heat calculations could be particularly relevant for materials with strongly anharmonic bonds (e.g., certain polymers, molecular crystals, or low-dimensional materials), where deviations from the Debye model have been reported. By fitting the experimental heat capacity data across various temperature regions, one can get the nonextensive parameter $q$, which can offer a phenomenological bridge between our theory and measurable quantities.

Our results establish a foundation for such investigations and demonstrate the value of combining nonextensive statistical mechanics with physically grounded quantum models like the Morse oscillator.
\section{Acknowledgement}
AG is deeply grateful to Dr. Shaon Sahoo for generously sharing his insights and offering valuable suggestions that helped shape this study.

\section{References}
\bibliographystyle{iopart-num}
\bibliography{manuscript}

\end{document}